\documentclass[referee]{aa} % for a referee version

\usepackage{graphicx}
\usepackage{natbib}
\usepackage{txfonts}
\begin{document}

\title{Small scale contributions to CMB: A coherent analysis}
\titlerunning{Small scale contributions to CMB: A coherent analysis}

   \author{Marian Douspis
\inst{1,2}
          \and
           Nabila Aghanim
\inst{2}
	  \and
	   Mathieu Langer
\inst{2}
         }

   \offprints{Marian Douspis}

   \institute{LATT, 14 avenue Edouard Belin, F-31400 Toulouse, France\\
              \email{douspis@ast.obs-mip.fr}
\and
Institut d'Astrophysique Spatiale (IAS), B\^atiment 121, F-91405 Orsay, France\\ 
Universit\'e Paris-Sud 11 and CNRS (UMR 8617)
}
            
%\date{\today}

%\begin{abstract}
\abstract{We reanalyse Cosmic Microwave Background data from
  experiments probing both large and small scales. We assume that
  measured anisotropies are due not only to primary fluctuations but
  also, especially at small scales, to secondary effects (namely the
  Sunyaev-Zel'dovich effect) and possible point source contaminations.
  We first consider primary and secondary anisotropies only.  For the
  first time in such analyses, the cosmological dependence of
  secondary fluctuations is fully taken into account. We show in that
  case that a higher value of the normalisation $\sigma_8$ is
  preferred, as found by previous studies, but also higher values of
  the optical depth $\tau$ and power spectrum index $n_s$ are needed.
  In the second part of our analysis, we further include possible
  contaminations from unresolved and unremoved point sources. Under
  these considerations, we discuss the effects on the cosmological
  parameters. We further obtain the best combination of relative
  contributions of the three kinds of sources to the measured
  microwave power on small scales at each frequency. Our method allows
  us to simultaneously obtain cosmological parameters and explain the
  so-called small scale power excess in a consistent way.
%\end{abstract}

%
%}

}

\maketitle

%
%________________________________________________________________

\section{Introduction}\label{intro}

Measurements of the Cosmic Microwave Background (CMB) angular power
spectrum are now becoming invaluable observables for cosmology. The
detailed shape of this spectrum allows one to determine cosmological
parameters with high precision. The CMB anisotropy spectrum has been
recently measured from large to intermediate scales with WMAP
\citep[][]{bennett2003}, Archeops \citep[][]{benoit2003} and BOOMERanG \citep[][]{B03}
experiments. At small scales, different experiments, ACBAR
\citep[][]{acbar}, CBI \citep[][]{cbi, cbi2}, VSA \citep[][]{rebolo2004} and BIMA
\citep[][]{bima}, have detected signal in excess of  that
expected from purely primordial CMB anisotropies. This excess is not
fully understood yet but several explanations have been proposed. Some
relate to the early Universe \citep[e.g.][]{cooray2002,griffiths2003}, 
some are associated with reionising sources
\citep[e.g.][]{oh2003} and others are related to the astrophysical
processes contributing to the signal at
the CMB frequencies \citep[e.g.][]{bond2005,toffolatti2005}.  As
a matter of fact, when observing at microwave wavelengths one expects
to detect not only primordial CMB fluctuations but also secondary
anisotropies as well as galactic and extra-galactic
foregrounds. Astrophysical contributions are, as much as possible,
removed during data analysis to produce clean estimates of the CMB
angular power spectrum. Nevertheless, residuals may remain and
contribute to the measured power.

In order to minimise galactic contamination, most of the small sky
coverage experiments (probing small scales) observe at high galactic
latitudes in regions where the sky is as much as possible free of
galactic emissions. Moreover, the galactic signal can be removed
through multi-frequency observations, or even by purely masking the
contaminated regions. Since most of the residual signal is expected on
large scales, galactic emissions only very moderately affect small
scale CMB observations.

At small angular scales, additional contribution to the primary CMB
signal arises from the interaction of CMB photons with matter along
their propagation path. These are the so-called secondary anisotropies
\citep[e.g.][ and references therein]{hu2002}. The latter are dominated
by the thermal Sunyaev-Zel'dovich (SZ) effect \citep[][]{sz72,sz80},
i.e. CMB photon inverse Compton scattering off free electrons in the hot
intra-cluster medium. When clusters are known and resolved by an
instrument, they can easily be cleaned out of CMB maps. However, not
all clusters
are known and we expect a contribution to the microwave signal from
unresolved/unknown clusters.  The SZ effect being frequency dependent
(contrary to primary CMB fluctuations) and its spectral signature being
known, one can use this information to detect and remove SZ signal
from unknown clusters in a map if multi-frequency observations are
conducted. On the contrary, with single frequency experiments, we are
consequently left with an SZ contamination at small scales.

Contamination by extra-galactic sources has long been studied for CMB
and SZ observations. The effects of radio sources were investigated
for low frequency experiments 
\citep[e.g.][]{ledlow96,cooray98,toffolatti98,sokasian2001,dezotti2005,gonzalez2005}.
At high frequencies dusty galaxies emitting in the infra-red are the
dominant sources of contamination
\citep[eg][]{blain98,white2003,lagache2005}.  Contamination by point
sources can be monitored by interferometric arrays, and multi-frequency
observations should help us in reducing the additional signal through
component separation.  Once again, however, a residual contamination is
likely to remain, especially in view of the uncertainties in the
modelling of the sources themselves.

The excess power reported by small scale CMB experiments may thus be
due to several astrophysical signals expected to contribute at those
scales. In order to correctly estimate the relative and possible
contribution of the secondary effects or extra-galactic signal on top
of the primary CMB anisotropy, one needs a coherent analysis.
Cosmological dependences are expected for CMB and SZ effect, and SZ
and point sources are contributing on rather similar scales. Thus,
unlike previous studies \citep[e.g.][]{goldstein2003,bond2005}, we do
not decorrelate small scales from those on large scales in our
analysis of the CMB data. On the contrary, we look for the best model
that fits data simultaneously on large and small scales. Hence,
assuming that primary, secondary and point sources signals are all
contributing to the model, we present a coherent analysis (i.e.
including the full cosmological dependence of secondary effects) of
the CMB angular power spectrum as measured from large ($\ell=2$) to
small ($\ell=10000$) scales.

In the next two sections we describe the CMB data we used and we
present our modelling of the power spectra of the different
contributions that enter the analysis. We then present our results in
terms of cosmological parameters inferred from an analysis accounting
coherently for primary, secondary and extra-galactic contributions. We
show in each case the relative contributions of all considered
signals, and discuss our results.

\section{Observations}\label{data}

CMB temperature anisotropies have been now measured on a wide range of
scales with high significance. WMAP has yielded the best measurements
from large to intermediate angular scales, and together with Archeops
and BOOMERanG, offers an estimate of a series of features (peaks and
troughs) consistent with those expected from acoustic oscillations in
the primordial fluid. ACBAR, CBI, VSA and BIMA observed smaller
scales, those affected by exponential damping of power of primary
fluctuations due to photon diffusion and to the finite thickness of
the surface of last scattering. The CMB polarisation (TE, EE, BB) has
also been measured by several experiments (\citep[DASI,][]{dasipol05},
\citep[BOOMERanG,][]{boomerang03_te, boomerang03_ee},
\citep[CBI,][]{cbipol04}, \citep[WMAP,][]{wmap_te}).  Altogether, CMB
temperature and polarisation anisotropy measurements now enable us to
constrain cosmological parameters and hence the underlying theoretical
models. A firm detection of the first peaks in the CMB anisotropy
angular power spectrum and a damping at smaller scales has now been
obtained. However, an excess of power at scales around a few
arcminutes ($\ell > 2000$) has been found in the data (see Sect.
\ref{intro}).

In the present study, we consider only a subset of data which allows
us to explore the whole range of measured scales. Namely, we use WMAP
and BOOMERanG data, whose major advantage is an accurate measurement
of the primordial signal at $\ell < 1500$, and that of ACBAR, BIMA and
CBI (at higher $\ell$) which exhibit an excess of power with respect
to the expected primary contribution. In addition, we take into
account in our analysis the polarised data from WMAP and BOOMERanG.
Altogether, this sample represents 1470 data points.

\begin{figure}[!h]
   \centering 
	\includegraphics[width=8cm]{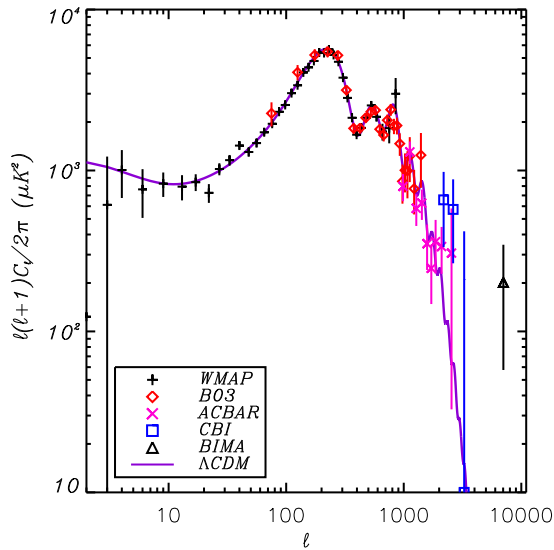} 
	\caption{Data used in the analysis. Overplotted the best model
	when only primary fluctuations are considered and fitted to
	data with $\ell < 1500$ (purple solid line).}  
	\label{powplotcmb}
\end{figure}

\section{Contributions to the signal}\label{contri}

Observations of the sky may contain many astrophysical processes
which contribute to the signal along the line of sight. They have to
be either accounted for or removed before analysing the signal of
interest. Usually CMB experiments observe high latitude patches of the
sky from which galactic contaminations are well subtracted. In the
following, we will therefore neglect potential galactic residuals and
focus exclusively on the extra-galactic contributions described below.

\subsection{Primary fluctuations}

We call primary fluctuations those anisotropies that are present at
recombination. They may be due to intrinsic temperature fluctuations,
density fluctuations or velocity fluctuations in the primordial
plasma. The derived CMB angular power spectrum shows a particular
shape the main characteristics of which are independent of the
cosmology: A plateau at large scales, a succession of peaks and a
damping at small scales. The details of those features are, on the
contrary, highly dependent on the cosmological parameters.
%(\sps
%Ajouter une ref? {\bf {\it MD: Maybe Wayne Hu }} ). 
The position of the first acoustic peak is sensitive for example to
the total density of the Universe. The CMB power spectrum is now well
used to determine cosmological parameters by fitting the data of
Fig.~1 with theoretical predictions. Here, the temperature and
polarisation angular power spectra for primary fluctuations were
computed using the Boltzmann code  \textsc{camb} \citep[][]{camb}.

\subsection{Secondary fluctuations - SZ effect}

The Sunyaev-Zel'dovich (SZ) effect is actually the dominant secondary
effect at scales between a few tens arcminutes to a fraction of
arcminute. It has now been observed in the direction of a few tens of
known clusters \citep[e.g.][ and references therein]{carlstrom2002}.
The SZ effect has a particular spectral signature which induces an
excess of brightness at high frequency and a deficit of brightness at
low frequency in the Rayleigh-Jeans regime. This makes it in theory
easily distinguishable from other contributions to the millimetre (mm)
and sub-millimetre (submm) sky when multi-frequency observations are
conducted. When the sky is observed at one single frequency
\citep[e.g. CBI, VSA, BIMA, AMIBA ][]{lo2001} it is \emph{a priori}
impossible to clean the signal from an SZ contamination except in the
direction of known clusters. An SZ contribution is thus expected at
small angular scales in CMB surveys. It is usually estimated through
the SZ angular power spectrum which can be inferred directly from
cosmological numerical simulations or computed analytically
\citep[e.g.][ for a review, and references therein]{komatsu2002}.

In the present study, we choose to compute the SZ power spectrum using
an analytic approach. Such a choice is particularly important for a
coherent analysis of the CMB anisotropies including primary and
secondary signals. Indeed, the SZ power spectrum depends on the
cosmological parameters through the number of SZ sources and their
characteristics (gas content, temperature, evolution...). Accounting
for modifications through numerical simulations requires to run
numerous simulations with different parameters which is both time and
storage consuming. Numerical simulations are also limited by
resolution and size. This affects the estimates of contributions at
small and large angular scales. The dependence on cosmological
parameters is more directly and easily handled with analytic
computations \citep[see for example][]{komatsu2002}. In practise, we
have computed the power spectrum of sources with masses between
$M_{\rm min}=5\,\times 10^{11}h^{-1}$ and $M_{\rm max}=5\,\times
10^{15}h^{-1}$ solar masses up to $z=7$ following
\citet[][]{komatsu2002}. The Poisson contribution to the SZ power spectrum
$C_\ell^{\rm SZ}$ is given by:
\begin{equation}
C_\ell^{\rm SZ}=g(\nu)^2\int_0^{z_{\rm
    max}}\,\frac{{\rm d}V(z)}{{\rm d}z}\,{\rm d}z\int_{M_{\rm
    min}}^{M_{\rm max}} \frac{{\rm d}N(M,z)}{{\rm
    d}M}[\hat{y}_\ell(M,z)]^2\,{\rm d}M
\label{eq:clsz}
\end{equation}
where $V(z)$ is the comoving volume element, ${\rm d}N(M,z)/{\rm d}M$
is the comoving mass function and $\hat{y}_\ell(M,z)$ is the 2D
Fourier transform of the projected Compton parameter (the line of
sight integral of the intra-cluster gas pressure) which characterises
the amplitude of the SZ effect. The spectral dependence is given by
$g(\nu)$.  We used a modified Press-Schechter mass function
\citep[][]{sheth99} and accounted for possible variations of the spectral
index $n_{\rm s}$ of the initial density power spectrum through the
``degenerated shape parameter'' defined in \citet[][]{wu2001}. For the
individual contribution $\hat{y}_\ell(M,z)$, we followed the
description in \citet[][]{komatsu2002} of a hot gas in hydrostatic
equilibrium in a universal dark matter potential with a constant
polytropic equation of state. Detailed effects of the gas profile and
mass function on the SZ power spectrum are given in \citet[][]{komatsu2002}
to which we refer the reader.  Using eq. (\ref{eq:clsz}) we can compute
the expected signal from a population of SZ sources for any given set
of cosmological parameters, at every multipole $\ell$ and for any
observing frequency $\nu$.  The SZ power spectrum has a bell-like
shape the amplitude of which depends mainly on the normalisation
parameter $\sigma_8$, but also on the mass and redshift ranges
together with the set of physical and cosmological parameters used in
the computation. A correlated contribution to eq. (\ref{eq:clsz})
would modify the low $\ell$ part of the spectrum by adding power at
large angular scales. The relative contribution of SZ anisotropies at
those scales being small we choose not to take the correlations into
account.  The EE and BB polarisation power spectra induced by clusters
are orders of magnitude smaller than the primary polarised power
spectra \citep[][]{liu2005}. We therefore do not take these into account in
the present analysis.

%\begin{figure}
%   \centering 
%	\includegraphics[width=8cm]{anal-simu-cl.ps} 
%	\caption{Comparison between our estimate of SZ power spectrum and the one derived from CLEF-SSH simulations}  
%	\label{fig1}
%\end{figure}

\subsection{Point source contamination}

The extra-galactic contaminations depend on the observing frequency
through the spectral energy distribution of the sources and their
number counts.  The extra-galactic contributions to the CMB signal are
usually classified broadly into radio galaxies emitting mainly at low
frequencies and dusty IR galaxies at high frequencies with a crossing
point at about 150 GHz where both contributions are of the same order
\citep[][]{ahl}.

Star forming galaxies emit, in the near-IR, optical and UV domains and
also in the far-IR (up to 90\% of their radiation) and submm, an
emission associated with the radiation absorbed and re-emitted by
dust. The total emission from unresolved and faint so-called IR
galaxies is responsible for the Cosmic Infra-red Background, the
fluctuations of which are now detected
\citep[]{lagache2000a,matsuhara2000,miville2002,kiss2001}. This is an
important contaminant for the SZ and CMB measurements at high
frequencies ($\ge 100$ GHz).

The radio galaxy emission is dominated by synchrotron emission from
relativistic electrons spiralling around magnetic field lines.  The
synchrotron spectrum generally follows a power law $\nu^\alpha$, with
$-1<\alpha<-0.5$. However, those spectra need not be simple power laws
\citep[][]{herbig92} but may have curved shapes instead; they even can
exhibit inverted slopes (with $\alpha>0$); moreover some of the radio
sources may be variable. Radio sources are rather well catalogued at
low frequencies thanks to NVSS \citep[][]{condon98} and FIRST
\citep[][]{white97} surveys. The all sky survey from WMAP
observations at 41 GHz \citep[][]{bennett2003} has provided us with
information at higher frequencies where the radio sources are poorly
known. In order to assess the level of contamination due to radio
sources at the frequencies of interest for the CMB (between 30 and 200
GHz), studies were conducted and models proposed to predict radio
source counts
\citep[][]{toffolatti98,sokasian2001,dezotti2005}. In particular, the
model by \citet[][]{dezotti2005} found to be in quite a good
agreement with detections of radio sources by WMAP provides us
with good estimates of the radio source contribution to CMB signal.

Point source contamination can be reduced in, or removed from, CMB
observations.  This is usually done using multi-frequency data and
component separation techniques, or using monitored observations at
the same frequency as CMB experiments, or finally using spatial
correlations with already existing source catalogues. Different
strategies for point source cleaning were used by the groups of ACBAR,
BIMA and CBI, and are described in detail in the articles by
\citet[][]{bima,mason2003,acbar} and
\citet[][]{cbi2}. These strategies comprise: Pointed observations
at the CMB frequencies of sources at lower frequencies, direct
counts using deep and mosaic images at the observing frequency, 
removal of faint sources from
external catalogues, ancillary surveys at lower frequencies, using
existing catalogues\dots  However, the uncertainties in the physical
description of the contaminating emissions (extrapolations from
frequency to frequency, source variability\dots) together with the
sensitivities of the complementary data leave an unremoved
contamination that is likely to contribute to the
signal. \citet[][]{toffolatti2005} have recently revisited the
contributions of extra-galactic point sources to the arcminute scale
measurements of ACBAR, BIMA and CBI at the relevant frequencies,
namely 28.5, 30 and 150 GHz. In particular using predictions by
\citet[][]{dezotti2005} updating the source model of \citet[][]{toffolatti98},
they have shown that faint sources with fluxes too weak to be detected
in ancillary surveys might contribute a significant fraction of the
arcminute signal.

For our study, we consider that such an unremoved component does indeed
contribute to the measured arcminute signal. We do not take into
account possible enhancement by lensing. We model the point source
contribution in a very simple way, assuming it is well described by
a power spectrum $C_\ell^{\rm PS}=\alpha_{\nu_{\rm obs}}^{\rm
PS}\ell^2$, typical of a Poisson distribution.  The parameter
$\alpha_{\nu_{\rm obs}}^{\rm PS}$ accounts for the {\it r.m.s.}
contribution of unremoved point sources (i.e. below the detection
limit or unaccounted for) at a given observing frequency $\nu_{\rm
obs}$. As a consequence, a different power spectrum, i.e. a different
$\alpha_{\nu_{\rm obs}}^{\rm PS}$ is assumed for each experiment (see
Sect. \ref{result}). Correlations between sources modify the
contribution at large angular scales by adding power to the shot noise
and thus they are expected to change the slope of the associated power
spectrum (\citep[e.g.][]{song2003,negrello2005}. However, in view
of the still limited knowledge we have of it, we choose not to take 
the source correlation into account.

Finally, point sources may in principle also contribute to the
polarised signal on small scales. Predictions on CMB polarisation
level from extra-galactic radio sources are summarised by \citet[][]{tucci}, 
who show that radio galaxy contribution is largely
dominated by synchrotron emission from our galaxy for all relevant
frequencies. Similarly, dusty IR galaxies are also expected to
contribute to the polarised signal. Not much data is available on IR
galaxy polarisation, and firm predictions on its contribution to the
CMB are difficult to make. Assuming the level of dust polarised
emission in external galaxies is not higher than that of the Milky
Way, it has been shown \citep[][]{tucci} that polarisation from IR
galaxies is sub-dominant at frequencies below 1000 GHz and for $\ell <
3000$.  We therefore neglect polarisation from radio and IR
galaxies in our analysis.

\section{Analysis}\label{result}

\subsection{Method}

In this work, we account for the contributions to CMB signal of the
different astrophysical sources (primary fluctuations, secondary
anisotropies and residual extra-galactic point sources) in a coherent
way. In other words, we compute their associated power spectra by
taking into account the dependences on the cosmological parameters of
each signal. Note however that our empirical modelling of residual
point sources does not explicitly exhibit such a dependence. We
analyse the CMB data shown in Fig.~1 and look for the set of best
cosmological parameters (best fit model) and corresponding error bars
when assuming one, two or three of the contributions described in
section \ref{contri} contribute to the measured signal.

As seen in the previous section, the SZ angular power spectrum is
frequency dependent. We thus consider three observing frequencies 150,
30 and 28.5 GHz.  The level of contamination from residual point
sources is related to the beam size of each experiment, to its
sensitivity and to the flux limit which accounts also for possible
source extraction using cross-correlations with other wavelengths.
Therefore, the point source contribution in terms of power spectrum is
fitted, in the following, independently for each experiment.  As a
result, the total power spectrum (sum of primary, secondary and
unremoved point sources) is also frequency dependent and thus fitted
for each observing frequency, i.e. for each small scale experiment
(ACBAR, BIMA, CBI) added to WMAP and BOOMERanG.

%{\bf NA: The following will be added if necessary only.}  This means
%that {\bf this is an approximation as the instruments are observing
%through a frequency band and not a single frequency. The signal should
%then be convoluted by the band to be a proper estimate of the
%contribution ... We have checked that the effect is negligible and
%keep the centre of frequency band approximation for simplicity and
%computation cost }

We use a Monte Carlo Markov Chain approach allowing us to account for
a large number of parameters without using intensive computational
resources. We apply the test of convergence developed by \citet[][]{joanna}
to check the robustness of our results. The likelihood of WMAP data is
computed by using the WMAP collaboration routines\footnote{These
routines are available at\\
http://lambda.gsfc.nasa.gov/product/map/current/m\_sw.cfm} whereas the
likelihoods of the other data (ACBAR, BIMA, CBI) are computed with the
Bond, Jaffe \& Knox approximation \citep[][]{bjk} when the necessary
information is provided (Gaussian-like approximation otherwise).  When
all the astrophysical contributions are taken into account the final
likelihood ${\mathcal{L}}$ writes:
$${\mathcal{L}} =  \prod_i^N {\mathcal{L}}_i(C_\ell(i, \nu_i, \theta) | C^{obs}_\ell(i))$$ 
where $C^{obs}_\ell(i)$ is the angular power spectrum of an experiment $i$
observing at frequency $\nu_i$, $\theta$ is the set of cosmological
parameters and the total theoretical angular power spectrum is:

$$C_\ell(i, \nu_i, \theta) = C_\ell^{\rm primary}(\theta) +
C_\ell^{\rm SZ}(\nu_i, \theta) + C_\ell^{\rm PS}(\nu_i),$$ 
with
$C_\ell^{\rm PS}(\nu_i)=\alpha_{\nu_{\rm obs}}^{\rm PS}\ell^2$.  Note
that in the case of WMAP and BOOMERanG experiments, the $C_\ell^{\rm
primary}(\theta)$ (and \emph{a fortiori} the likelihood $\mathcal{L}_i$)
includes polarisation signal.  We consider throughout a flat
cosmological model without contribution from neutrinos nor tensor
modes. We assume a power law initial power spectrum for the density
perturbations with spectral index $n_s$. The remaining cosmological
parameters are: The baryon budget, $\Omega_{\rm b}h^2$, the cold dark matter
content, $\Omega_{\rm cdm}h^2$, the Hubble constant, $H_0$, the
normalisation factor  $\sigma_8$ (given by the amplitude of fluctuation on a
sphere of 8$h^{-1}$ Mpc), and the Thomson scattering
optical depth $\tau$ (related to the reionisation history of the
Universe), to which we add the amplitudes, $\alpha_{\nu_i}^{\rm PS}$, of the point source
contributions for each experiment.

\subsection{Results}

In the following we consider first the case where primary and
secondary anisotropies are present, then we add the contribution of
unremoved point sources. We show in both cases the set of derived
cosmological parameters and point out the effects of an additional,
non primary contribution to the signal.

\subsubsection{CMB+SZ case}\label{cmb+sz}

Atmospheric, galactic and resolved point sources are usually well
removed from CMB observations; SZ effect has thus firstly been
advocated to explain the excess of power detected at small scales.
Previous analyses often decorrelate anisotropies at large scales,
associated with primary signal, from those at small scales related to
secondary signal. A primary power spectrum is thus generally fitted
with data at $\ell < 1500$. It is added to a template SZ power
spectrum (generally obtained from numerical simulations with fixed
cosmological parameters). In such analyses, the amplitude of the SZ
power spectrum is the only parameter set free (often labelled
$\sigma_8^{SZ}$ and related to the normalisation $\sigma_8$). Such a
template spectrum is then fitted to the data available above
$\ell=1500$ and the SZ contribution, and therefore $\sigma_8^{SZ}$, are
derived.

As seen in Sect. \ref{contri}, the SZ angular power spectrum depends
on the cosmological parameters through the volume element, the number
density and the individual Compton parameter. The relative
contribution of primary and secondary signals are thus strongly
connected and cannot be formally separated. As a result, although the
signal from primary fluctuations is weak at small scales ($\ell>1500$)
it is incorrect to decorrelate the two contributions in the fit.

\begin{figure}
   \centering \includegraphics[width=8cm]{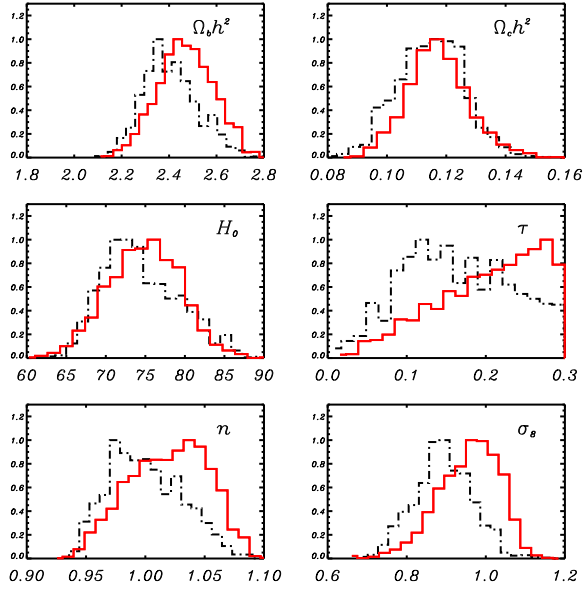}
   \caption{Marginalised one-dimensional distributions for the
     cosmological parameters investigated. The black dotted-dashed
     line shows the curves for the case where data with $\ell < 1500$
     are fitted with pure CMB models. The red solid lines are derived
     from the MCMC run with all data and SZ contribution added to CMB
     (Run 1). The value of $\Omega_bh^2$ as been multiplied by 100 for
     clarity. }
   \label{mcmc1}
\end{figure}

In the present study, we thus fit all data described in Sect.
\ref{data} by angular power spectra summing the primary and secondary
contributions, with free cosmological parameters. The results of the
corresponding MCMC run (Run 1) are shown in terms of one dimensional
likelihood functions (solid red lines) in Fig.~\ref{mcmc1}. The latter
are compared with the likelihood functions obtained by fitting $\ell
<1500$ data by primary CMB power spectra only (black solid lines).

\begin{figure}
   \centering \includegraphics[width=8cm]{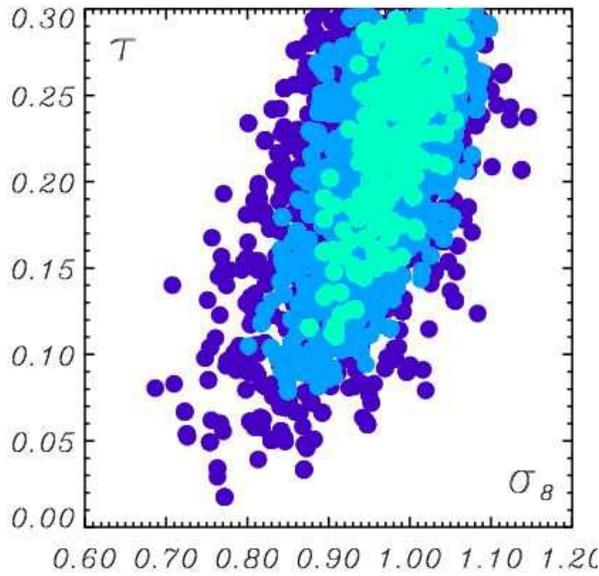}
   \caption{Two-dimensional plot of the degeneracy between $\sigma_8$
   and $\tau$ obtained from Run 1. Coloured shaded regions mark the 1,
   2, 3 $\sigma$ contours from light to dark} \label{degntau}
\end{figure}

The main effect of adding coherently the SZ signal is to allow the
cosmological models to fit the small scale data. This implies that
high values of $\sigma_8$ (median value: 0.97) and $n_s$ (median
value: 1.02) are preferred. Both parameters tend to amplify the
power at small scale. In particular, as seen before, the amplitude of
the SZ power spectrum depends on $\sigma_8$. The higher $\sigma_8$ the
larger the SZ signal. Such an effect was recognised in previous
studies \citep[][]{goldstein2003,bond2005}. A coherent analysis taking
into account both primary and SZ signals allows to exhibit the effects
of a small scale excess of power on the other cosmological parameters.
As a matter of fact, in order to compensate the additional small scale
power and keep the large scale power at WMAP amplitude, large values
of the optical depth $\tau$ (close to 0.3) are necessary to balance
the effects of $\sigma_8$ and $n_s$. Such values are indeed allowed by
the WMAP TE power spectrum at large scales \citep[see][]{wmap_te,
  melch2005}.  This is the well known degeneracy between $\sigma_8$ or
$n_s$ and $\tau$ illustrated in Fig.~\ref{degntau}. The other
cosmological parameters $\Omega_{\rm b}h^2$, $\Omega_{\rm m}h^2$ and
$H_0$ are only very marginally affected by the excess of power at
small scales, but our results seem to suggest that slightly higher
values of $\Omega_{\rm m}h^2$ and $H_0$ are favoured.

\begin{figure}
   \centering \includegraphics[width=8cm]{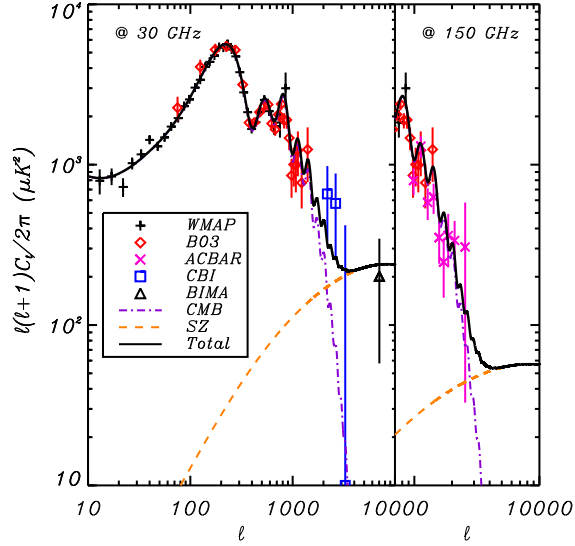}
   \caption{Total power spectrum (in solid black line) overplotted to
     the data for the two frequencies considered. The SZ power
     spectrum amplitudes are function of the frequency range
     considered (magenta dashed lines) while primary (purple
     dot-dashed line) contribution is not.}
   \label{powplotcmbsz}
\end{figure}

\begin{figure}
   \centering \includegraphics[width=8cm]{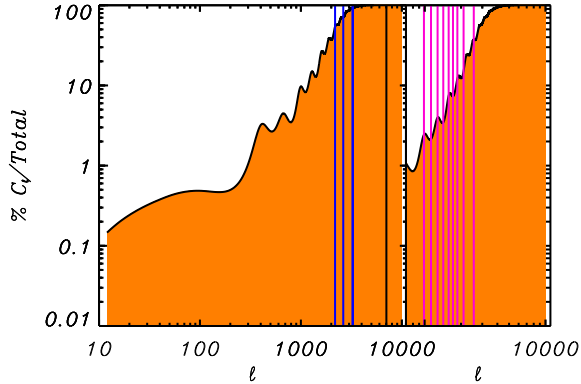}
   \caption{Contributions of primary and SZ power spectra to the total
     spectrum (solid black line of Fig.~\ref{powplotcmbsz}). The
     orange shaded (white) region gives the percentage of total power
     due to SZ (resp. primary fluctuations) contribution.}
   \label{contribcmbsz}
\end{figure}

The best model, defined as the one with cosmological parameters being
the median of the distributions (Fig.~\ref{mcmc1}), is shown in
Fig.~\ref{powplotcmbsz}. The corresponding cosmological parameters
are: $\Omega_bh^2= 0.025$, $\Omega_{cdm}h^2 =0.116$, $H_0=75.33$
km/s/Mpc, $\tau=0.22$, $n=1.02$, $\sigma_8=0.97$.  This model slightly
improves the goodness of fit as compared to the best pure primary
spectrum without adding any other parameter. In
Fig.~\ref{powplotcmbsz}, the corresponding total angular power
spectrum is plotted in black solid line whereas the primary and
secondary contributions appear respectively in purple dotted-dashed
line and orange long-dashed line.  The difference in SZ amplitude
between the two plots at 30 and 150 GHz is due to the frequency
dependence $g(\nu)^2$ in eq. (\ref{eq:clsz}). The relative
contributions to the total power spectrum are shown as functions of
$\ell$ in Fig.~\ref{contribcmbsz}; the primary contribution is
displayed in white and the secondary contribution is displayed in
orange/grey. This figure gives the contribution of the SZ signal
observed in microwave in a joint and coherent analysis where
cosmological parameters are adjusted simultaneously for both primary
and secondary anisotropies to fit current data.  At $\ell = 3000$ the
CMB signal is dominated by the SZ contribution which represents 86\%
and 59\%, at 30 and 150 GHz respectively, of the total signal.
%Such contributions could be as low
%as ?? and as high as ??.  {\bf {\it MD: I
%need a new run, trying to launch it today (09/01). I am not convinced
%yet}} 
As one can see in Fig~\ref{mcmc1}, the best model that fits  $\ell
< 1500$ data, free from SZ contribution, is still a good model for the
whole data set (up to $\ell = 10000$). As a matter of fact, the SZ
contribution at 30 (150) GHz at $\ell =3000$ varies from
0 to 98\% (91\%) for the models within the 68\% confidence interval in
the 6 dimensional cosmological parameter space. The actual data set
at small scale is clearly poorly constraining due to the large error
bars and the inter--calibration problem. The situation is then expected to
improve with future observations. Furthermore, it is worth noting that
the present analysis does not account for the statistical dispersion
which induces a large dispersion in the SZ contribution. Most of
the SZ signal is indeed due to rare massive clusters. As noticed in
Fig. 2 of \citet[][]{cooray2002}, the SZ variance is expected to be
much larger due to the non-Gaussian nature of the SZ signal than that
of a Gaussian distribution.
%It is twice as large around
%$\ell=3000$. Such an effect is even stronger for small scale coverage
%experiment. 
%The SZ contribution at $\ell=3000$ could therefore be a
%factor two smaller than what we derived from
%Fig.~\ref{contribcmbsz}. 
Mutli-frequency observations of the same field will then be the only
way around this problem.

\subsubsection{CMB+SZ+PS case}\label{cmb+sz+ps}

As seen in Fig.~\ref{powplotcmbsz}, the SZ angular power spectrum does
not fully explain the excess power of some data points at small
scales. Of course, the latter could be 1$\sigma$ events. Nevertheless,
as discussed above, some contribution from unresolved and unremoved
point sources is also expected in the small scale signal. We want to
quantify the effects of this additional contribution.  We thus repeat
our analysis by including the point sources in addition to primary and
SZ contributions (MCMC Run 2).  Modelling the angular power spectrum
by $\ell(\ell+1)C_\ell/2/\pi \equiv \alpha_{\nu_i}^{\rm PS}
(\ell/2000)^2$, we introduce 3 new parameters $\alpha_{\nu_i}^{\rm
  PS}$ noted $A,B,C$ and describing the amplitudes of point sources
remaining in each experiment, ACBAR, BIMA and CBI respectively.  These
new parameters give the amplitude in $\mu \rm K^2$ of the point source
contribution at $\ell=2000$.

\begin{figure}[!h]
   \centering \includegraphics[width=8cm]{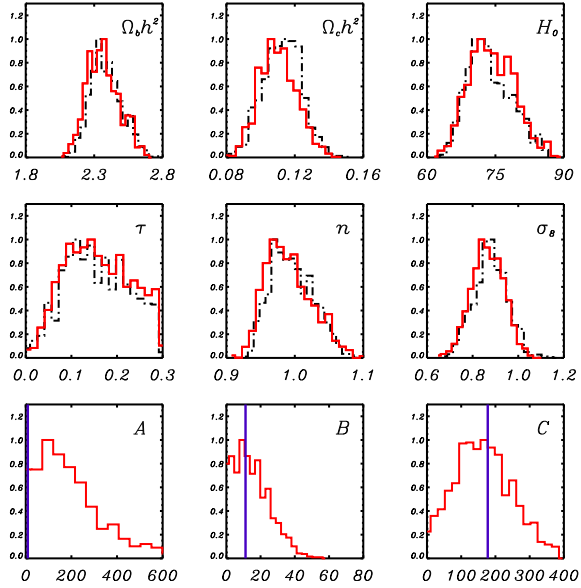}
   \caption{One-dimensional distribution of cosmological and point
     sources amplitude parameters from MCMC Run 2.  The vertical
     (blue) lines in the three lower panels give the amplitude of
     point source contributions as computed by Toffolatti et al. (see
     text).} \label{mcmc2}
\end{figure}

The output of Run 2 is shown in Fig.~\ref{mcmc2} again in terms of one
dimensional likelihood functions. We notice in this case that the
estimated cosmological parameters (red/light lines) are all in perfect
agreement with the results obtained by considering $\ell < 1500$ data
and primary fluctuations only (black/dark lines). The addition of an
unremoved point source contribution at small scales improves the
goodness of fit more than when adding SZ only. The values of
cosmological parameters $\Omega_{\rm b}h^2$, $\Omega_{\rm m}h^2$ and
$H_0$ are almost identical to those obtained with primary CMB only and
$\ell < 1500$ data. The parameters $\tau$, $n_s$ and $\sigma_8$ also
converge to values obtained in pure primary CMB analyses. In the
present case (primary CMB, SZ effect and unresolved point sources) we
find a perfect agreement between large and small scale data converging
towards the following median values for the cosmological parameters:
$\Omega_bh^2= 0.0237$, $\Omega_{cdm}h^2 =0.111$, $H_0=74.6$ km/s/Mpc,
$\tau=0.16$, $n=0.99$, $\sigma_8=0.87$.

In addition to the cosmological parameters, we furthermore obtain
estimations of the point source amplitudes for each experiment. The
corresponding values are 181, 14, 170 $\mu \rm K^2$ for ACBAR, BIMA,
CBI respectively. A recent study based on a physical model for point
sources in radio and infra-red domains \citep[][]{toffolatti2005} has
shown that the excess of power at small angular scales could be
explained almost entirely by the contribution from unresolved point
sources. We compare the predictions of \citet[][]{toffolatti2005} to
our results given by the one dimensional likelihood functions of $A,
B$ and $C$ in Fig. \ref{mcmc2}.  We find that predictions from
Toffolatti et al. are not only in agreement with our results but their
values correspond to the maxima of the distributions in Fig.
\ref{mcmc2}.

\begin{figure}%[!th]
   \centering \includegraphics[width=8cm]{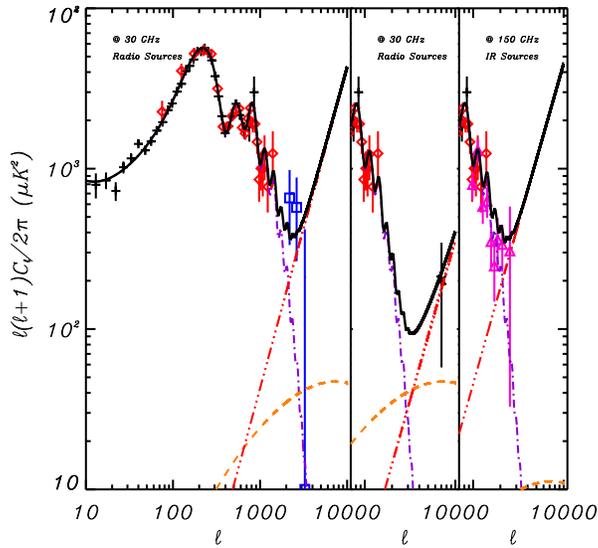} \caption{Best
   model using all contributions. Primary in purple, SZ in orange,
   point sources in red.} \label{powplotcmbszps}
\end{figure}

\begin{figure}
   \centering \includegraphics[width=8cm]{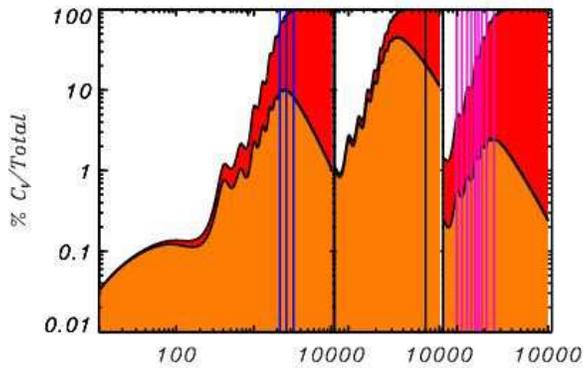}
   \caption{Relative contributions of primary CMB, SZ effect and
   unremoved point sources power spectra. Primary CMB is displayed in
   white, SZ effect is in orange/light grey and point sources are in
   red/dark grey. From left to right, the vertical lines mark the
   $\ell$s observed by CBI, BIMA, ACBAR experiments.
%(\sps C'est dans quel ordre? Acbar, Bima, Cbi?)
}
	\label{contribcmbszps}
\end{figure}

We plot the contributions of each component to the signal in
Fig.~\ref{powplotcmbszps} for our best fit model. The relative
contributions of the three astrophysical processes to the total signal
are summarised in Fig.~\ref{contribcmbszps}.  The primary CMB is
displayed in white, the SZ effect is in orange/light grey and the
point sources are in red/dark grey. At $\ell= 3000$, the relative
contributions are 2, 40 and 9\% for SZ contributions at (ACBAR, BIMA,
CBI frequencies) and 91, 32 and 85\% for the point source signals (at
the same frequencies).  Again, these numbers, corresponding to the
best model, have to be taken with caution as they do not account for
the larger statistical dispersions due to the non-Gaussian nature of
both SZ and point sources. Furthermore when the 68\% confidence
interval in 6 dimensions is considered, the SZ (PS) contributions at
$\ell=3000$ vary from 0 to 80\%, 0 to 96\% and 0 to 91\% (0 to 99\%, 0
to 85\% and 0 to 98\%) for respectively ACBAR, BIMA and CBI.
 
%{\bf {\it MD: Something about the errors ?? numbers ?? 
%rappel: contrib SZ: 0 a 70\%, 0 a 94\%, 0 a 82\% contrib CMB: 1 a
%90\%, 4 a 87\%, 2 a 56 \% for acbar, bima, cbi resp.}}

\section{Discussion}

As we have seen in Sect. \ref{cmb+sz}, the excess power on small
scales could be attributed to the sole contribution of SZ sources
(86\% and 59\% at 30 and 150 GHz at $\ell =3000$). In that case, high
values of $\sigma_8$, $n_s$ and of the Thomson scattering optical
depth $\tau$ are favoured. This in turn may suggest that the Universe
went through a very extended reionisation period, which remains
compatible with present observational constraints as reported by
\citet[][]{melch2005}. However, once we include the possibility of
unremoved or unresolved point source contribution, the likelihood
functions of $\tau$ and $\sigma_8$ that we obtain peak essentially at
the same values as obtained with $\ell < 1500$ primary CMB data only.

Our method parametrising, through the parameters $A, B, C$, the
extra-galactic source power spectrum enables us to probe the possible
contributions to the signal. Furthermore, it allows to test the
consistency of specific PS physical models with the cosmological
model. In case of a mismatch, our method would imply the need for
different physical descriptions of the source (number counts,
evolution, spectral energy distribution).

A substantial fraction of point source contribution seems 
allowed (see Fig. \ref{contribcmbszps} and Sect. \ref{cmb+sz+ps}).
Such a contribution is in agreement with extra-galactic number count
predictions \citep[][]{toffolatti2005}.  We noted however that
statistical dispersion due to the non-Gaussian nature of SZ and point
sources is likely to reduce significantly the actual contributions. In
addition given the current data, large ranges of PS contribution are
allowed. From the CMB standpoint, despite the excellent agreement of
Toffolatti's model, on the one hand, the SZ effect could effectively
contribute and solely explain the excess of power at small scales. On
the other hand, different point source models are allowed by current
data.

Despite their quality, current data are still not constraining enough
for such kind of multi-contribution studies. In addition to their
intrinsic error bars, the use of multiple experiments introduces
additional inter-calibration uncertainties. Ideally, we would need an
experiment that would observe continuously from $\ell =2$ to
$\sim 10000$. The Planck satellite will only observe up to $\ell \sim
2000$. However, we will still lack data at high and intermediate
multipoles, especially in the range $\ell \in [2000, 10000]$. In a few
years from now, a combination of Planck and future ground based high
resolution experiments will help us in reducing significantly the
error bars and solving partly for the inter-calibration problem.

%Future study including residuals from SZ effect (cuts in the mass
%redshift space) and from the Galaxy. application for Planck and SPT.

%Limitations of our present study: observational limitations (big error
%bars and calibration uncertainties), theoretical limitations (gas
%distribution in clusters, cosmological parameters to take into account
%like dn/dk, physical model of sources).

\section{Conclusions}

We presented the first coherent analysis of CMB data available from
$\ell =2$ to $\ell = 10000$, taking simultaneously into account
contributions from the primary fluctuations and secondary anisotropies
as well as point source contamination. More specifically, we have
explicitely taken into account in a consistent way the cosmological
dependence of secondary anisotropies. In the first part of our work,
this allowed us to confirm that SZ effect contribution to the CMB
power spectrum could explain the so-called small scale power excess,
as found in previous studies. We showed however that this requires not
only high values of $\sigma_8$, but also high values of the spectral
index $n_s$ and of the Thomson scattering optical depth $\tau$,
leaving other cosmological parameters basically unaffected.

In the second part of our work, we included also unresolved/unremoved
point source contamination. In that case, we recover the cosmological
parameter values as determined from large scale CMB data only.  Our
consistent analysis introduces a parametrisation of the point source
signal. Consequently, we obtain estimates of the contribution of each
CMB source to the total power spectrum. Our method hence allows us to
test for cosmological consistency of physical models of point source
populations.

\begin{acknowledgements}
The authors thank Simon Prunet for discussions. MD acknowledges
financial support from CNES. 
      
\end{acknowledgements}

\bibliographystyle{aa}
\bibliography{dal_RC}

\end{document}